# A Three Phase Semantic Web Matchmaker

Golsa Heidary
*Young Researchers Club,*
*Islamic Azad University,*
*Najafabad Branch,*
*Isfahan, Iran.*
golsa_heidary@sco.iaun.ac.ir

Kamran Zamanifar[1],
Naser nematbakhsh[2]
*Dept. of Computer Science,*
*Islamic Azad University,*
*Najafabad Branch, Isfahan, Iran.*
{[1]zamanifar,[2]nemat}@eng.ui.ac.ir

### Abstract

*Since using environments that are made according to the service oriented architecture, we have more effective and dynamic applications. Semantic matchmaking process is finding valuable service candidates for substitution. It is a very important aspect of using semantic Web Services.*

*Our proposed matchmaker algorithm performs semantic matching of Web Services on the basis of input and output descriptions of semantic Web Services matching. This technique takes advantages from a graph structure and flow networks. Our novel approach is assigning matchmaking scores to semantics of the inputs and outputs parameters and their types. It makes a flow network in which the weights of the edges are these scores, using Ford-Fulkerson algorithm, we find matching rate of two web services. So, all services should be described in the same Ontology Web Language. Among these candidates, best one is chosen for substitution in the case of an execution failure. Our approach uses the algorithm that has the least running time among all others that can be used for bipartite matching.*

*The importance of problem is that in real systems, many fundamental problems will occur by late answering. So system`s service should always be on and if one of them crashes, it would be replaced fast. Semantic web matchmaker eases this process.*

## 1. Introduction

Semantic web service matchmaking is the process of finding an existing Web services based on the description of their functional and nonfunctional semantics [6]. Matchmaking scenarios typically occur when one is trying to reuse an existing piece of functionality (represented as a Web service) in building new or enhanced business processes. Central to the majority of contemporary approaches to Semantic Web service selection is that the functionality of Web services is logically defined in, for example, the standard first-order description logic-based ontology language OWL [16] or a rule language like SWRL, or a logic programming language like F-Logic. In any case, intelligent agents can exploit standard means of logic reasoning to automatically understand the Web service semantics, in particular to determine the degree to which the service is semantically relevant to a given service request.

Nowadays, academic as well as industrial communities focus one part of their researches on Web services technology like Web services matching. The basic architecture of Web services implements a Service Oriented Architecture (SOA), which allows their integration to internal or external applications. Service oriented architectures are becoming popular since they provide more effective and dynamic applications [31]. A Web service is a software system designed for interacting interoperable machines via the internet. They are based on eXtensible Markup language (XML) which constitutes the basic technology of Web services. Web services are, usually syntactically, described with standards like UDDI, SOAP and WSDL. Using semantic Web Services in service oriented architectures improves interoperability and scalability. A transaction with a service requires at least two parts: the service requester seeking a service to complete its work, and the service provider providing a service sought by the user. Semantic matchmaking is the process of finding suitable Web Services that satisfies the request. As the number of available Web Services on the Internet increases, finding a suitable Web service satisfying the needs becomes more difficult.

Universal Description, Discovery and Integration (UDDI) is a virtual registry that exposes information about Web services [10].

Simple Object Access Protocol (SOAP) is a protocol to exchange structured information in distributed environments [32]. It uses XML to define an extensible framework of messages which provides a constructed message that can be exchanged through a variety of underlying protocols. The protocol SOAP is independent from any particular programming model and from any specific semantics of an implementation [2].

Web Service Description Language (WSDL) provides a model and an XML format for describing the Web services. It separates the description of the abstract functionality, offered by a service, from the concrete details of a service description such as "how" and "where" [6]. WSDL describes only the syntactic interface of Web services. Hence, the pure WSDL cannot be used for automatic Web services matchmaking: Semantics are required in order to make information accessible to agents. The purpose of this work is to present, in one hand, a model of semantic Web service matchmaking. Also the publication of a service in UDDI doesn`t allows the semantic matching of Web services. Then to go beyond these limits, we propose a manner of Web services storage in which all the existing services are in OWL-S [13], which provides a high level description of the services capabilities. OWL-S provides service profile class which includes IO (input, output) capabilities of the services. During the matchmaking process, requests and advertisements (service offers) described as OWL-S documents are matched according to IO capabilities and the best advertisement is selected or a list of matching results is generated to make the choice manually. The difficulty in the matchmaking process occurs when there is no exact match for the request. Partial matches must be evaluated in these situations. So, a matchmaker needs to determine both exact and partial matches in a comparable fashion. IO attributes describe syntactically which inputs are required by a service to function, and which outputs are returned [24].

The obtained algorithm of matchmaking draws its advantages from a flow network and also from a semantic annotations and similarity measures between parameters. The introduction of semantics in the description of web resources reflects new achievements in web services technologies, through extensive specifications, automation of services selection, composition and translation of message content, self-describing service monitoring and recovery from failure [21]. Semantic web services assure more machine-oriented expressive power and usage of services, completely transforming the web information access from the usual content-based retrieval to semantic annotated functionalities, exposed by the web services.

The goal is to improve the mediation activity among service providers through proactive integration for providing automated semantic interoperability. The approach exploits the agent paradigm for achieving matchmaking.

The paper is organized as follows: Section 2 is devoted to the related work concerning semantic annotations and the synthesis of matchmaking. In Section 3, we present the proposed approach, its implementation and its algorithms and explane the three phases of our work completely. Section 4 summarizes the algorithm's complexity and the advantages of our approach compared to the other approaches. And finally in Section 5, we end with concluding remarks and future works.

## 2. Related work

The computation methods for the similarity between Web services are studied and applied in many aspects. Woogle, the Web service search engine, supports similarity search for Web services. Keyword search paradigm is insufficient for Web service search in that the underlying semantics cannot be captured. Motivated by the above fact, the technique to support similarity search for Web services was proposed by [6]. In this approach, similar operations are determined mainly by using the association-rule-mining approach and a hierarchical clustering algorithm for parameter names of Web services. However, the association rules are not very effective when the associations among services are complex, and it is difficult to fully represent the causal relationships implied among them. Further, the reasoning, critical for automatic matchmaking and discovery of Web services, can`t be done straightforwardly. Similarities between Web services are applied to obtaining Web services communities. [23] Proposes the nearest-neighbor approach to obtain the classes for the given services. The similarity measure just considers whether two services are similar, but does not explore how similar they are.

The ontology based modeling gives semantic models as conceptual frameworks for the semantic description of Web services, in which the ontologies were regarded as the semantic annotations [17].

In many applications, the strictly numeric representations will always have to sacrifice efficiency due to the inappropriate preciseness, such as the Web service search in which requesters always want to locate the desired ones as soon as possible [7].

Ontologies are used in order to incorporate semantics in web service descriptions. An Ontology models domain knowledge in terms of Concepts and Relationships between them. OWL [15] has evolved as a standard for representation of ontologies on the Web. OWL-S [16], formerly called DAML-S [12], defines ontology for semantic web services. OWL-S describes a service in terms of its Service Profile, Process Model and Grounding. The Service Profile models the Inputs, Outputs, Pre-conditions and Effects (IOPE) of the service [1]. The Inputs and Outputs in the Service Profile refer to concepts in ontologies published on the web. Service advertisements and search queries are both expressed in terms of OWL-S descriptions. An ontology reasoner is an important component in the process of semantic matchmaking. A reasoner can infer additional information which has not been explicitly stated on ontology. Subsumption, concept satisfaction, equivalence and disjointedness are some examples of reasoning operations. Many of these operations are used in the semantic matchmaking process.

The Service Profile contains enough information for a matchmaker to determine if a service satisfies the requirements of a client. In fact, several matchmaking algorithms rely only on the matching of Inputs and Outputs of the Service Profiles [19].

Generally, it is important to develop a semantic model to describe the inherent causal relationships among the given Web services. To obtain the semantics implied by the services themselves, mining the historical invocations and behaviors is doomed. To facilitate automatic matchmaking, developing an efficient measure for semantic associations among multiple Web services is indispensable.

## 3. Finding matching rate of two web services

Our matchmaker works in three phases.

In the first phase, it compares two web services` input/output parameters, semantically. Since all advertised services are described in the same Ontology Web Language (OWL), so we can easily compare the capability and functionality of them. The result of this phase is PARSIM.

In the second phase, we compare the type of parameters of input/output of two web services. The result of this phase is PARTYPE.

In the third phase, with both PARSIM and PARTYPE we compute the matching rate of two semantic web services.

### 3.1. First phase

The first phase is computing the similarity rate of parameters of two services` input/output. We want to compare two web services` functionalities. So we should compare inputs and outputs with each other, individually. The comparing process for both inputs/outputs is the same. So we explain the output comparing.

Assume that we want to compute the matching rate of services A and B. Each service`s input is shown by $A_{in}$ and $B_{in}$, respectively and the outputs are shown by $A_{out}$ and $B_{out}$. So we compare $A_{in}$ with $B_{in}$ and $A_{out}$ with $B_{out}$. In the OWL, we have many classes of words. When we compare two words with each other, they may be in one class, in different classes or one is in the super class of the other one. According to these situations, we give scores to them. Therefore, four results will be achieved. Algorithm 1 by the name of CASE, shows the scoring method. The inputs of this algorithm are output parameters of two web services that are matching. If the algorithm returns 10, it means that these two parameters can be replaced with each other. If it returns 0, the matching is failed and these parameters can`t be used instead of each other. Score 7 shows that the parameters to some extent match. So the decision making will be after type matchmaking. Score 3 is not good and the substitution of a service will depend on other parameters matching rate and type.

An Equivalent algorithm also is used for inputs.

| *Algorithm 1 : Case ($A_{out}$, $B_{out}$):* |
|---|
| *1 :  If ($A_{out}$ and $B_{out}$ are in the same class) then* |
| *2 :      return 10;* |
| *3 :  If ($A_{out}$ is subclass of $B_{out}$) then* |

```
4 :    return 10;
5 : If (B_out subsumes A_out) then
6 :    return 7;
7 : If (A_out subsumes B_out) then
8 :    return 3;
9 : Otherwise
10 :   return 0;
```

**3.1.1. Steps of matchmaking:** We do the matching by the help of bipartite graph. So in the first step of phase one, we make a bipartite graph. A Bipartite Graph is a graph $G = (V, E)$ in which the vertices set can be partitioned into two disjoint sets, $V = V_0 + V_1$, such that every edge e in E has one vertex in $V_0$ and another in $V_1$. The matching is complete if and only if, all vertices in $V_0$ are matched. It means that all vertices in $V_0$, as well as $V_1$, should have an edge.

Let $A_{out}$ and $B_{out}$ be the set of output concepts in A and B respectively. These constitute the two vertex sets of our bipartite graph. Construct graph $G = (V_0 + V1, E)$, where, $V_0 = A_{out}$ and $V_1 = B_{out}$. Consider two concepts a in $A_{out}$ and b in $B_{out}$. It means that a is one of the output parameters of A and b is one of the output parameters of B. Let R be the result of CASE (in our algorithm, which can be 10, 7, 3, 0) between concepts a and d. We define an edge (a, d) in the graph and label this edge as R. Therefore if matching is complete (all vertices have at least one edge), now we compute the whole matching rate for these two services.

In "figure. 1" we have an example of a bipartite graph which has the complete matching.

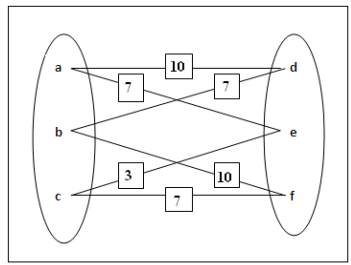

**"Figure 1. An example of bipartite graph of output concepts"**

In the second step, we must use an algorithm for computing matching rate. It will be done by the help of flow networks. Therefore at first, we will explaine flow networks, in brief.

*What is flow network?* To make a flow network, we can use a directed graph and use it to answer questions about flows of materials. Assume that a material coursing through a system from the source to the sink. In a source, the material is producing and in a sink it is consumed. The source produces the material at some steady rate, and the sink consumes the material at the same rate. Each directed edge in a flow network can be thought of as the similarity rate. Each edge has a definite capacity which is given as a maximum rate. Vertices (in our usage) are the input or output parameters of two web services which are comparing with each other than the source and sink. It means that we should add the source and sink vertices to our bipartite matching graph. In the maximum-flow problem, we will to compute the greatest rate at which material ships from the source to the sink without violating any capacity constraints.

This problem can be solved by efficient algorithms. There are two general methods for solving the maximum-flow problem which are Ford-Fulkerson and Edmonds-Karp, and some others which are push-relabel, relabel-to-front, Hopcroft-Karp bipartite matching algorithms. For finding the matching rate of two semantic web services, we use the first method that we will describe it in the following.

*Max flow in a Flow network:* A flow network G = (V, E) is a directed graph in which each edge (u, v) $\in$ E has a nonnegative capacity c (u, v) $\geq$ 0, unless we assume that c(u, v)= 0. We have two vertices in a flow network: a source s and a sink t. For convenience, we assume that every vertex lies on some path from the source to the sink. That is, for every vertex     v $\in$ V, there is a path s; v; t. So the graph is connected and $|E| \geq |V| - 1$.

Each flow network has three properties:
- Capacity constraint: For all u, v $\in$ V, we require f (u, v) $\leq$ c(u, v).
- Skew symmetry: For all u, v $\in$ V, we require f (u, v) $= -$ f (v, u).
- Flow conservation: For all u $\in$ V $-$ {s, t}, we require

*Residual networks:* Given a flow network and a flow, the residual network consists of edges that can admit more flow.

Suppose that we have a flow network G = (V, E) with source s and sink t. Let f be a flow in G, and consider a pair of vertices u, v $\in$ V. The amount of additional flow we can push from u to v before exceeding the capacity c (u, v) is the residual capacity of (u, v), given by

$$c_f(u, v) = c(u, v) - f(u, v) \qquad (1)$$
$$E_f = \{(u, v) \in V \times V : c\, f(u, v) > 0\} \qquad (2)$$

That is, as promised above, each edge of the residual network, or residual edge, can admit a flow that is greater than 0.  The edges in $E_f$ are either edges in E or their reversals.

If f (u, v) < c (u, v) for an edge (u, v) $\in$ E, then $c_f$(u, v) = c(u, v) $-$ f(u, v) >0 and (u, v)$\in$ $E_f$. If f (u, v) > 0 for an edge (u, v) $\in$ E, then f (v, u) < 0. In this case, $c_f$(v, u) = c(v, u)$-$f(v, u) >0, and so (v, u) $\in$ $E_f$. If neither (u, v) nor (v, u) appears in the original network, then:
c (u, v) = c (v, u) = 0, f (u, v) = f (v, u) = 0 and $c_f$ (u, v) = $c_f$ (v, u) = 0.

We conclude that an edge (u, v) can appear in a residual network only if at least one of (u,v) and (v, u) appears in the original network.

The residual network $G_f$ is itself a flow network with capacities given by $c_f$.

*Augmenting Paths:* In a flow network G = (V, E) and a flow f, an augmenting path p is a simple path from s to t in the residual network $G_f$. By the definition of the residual network, each edge (u, v) on an augmenting path admits some additional positive flow from u to v without violating the capacity constraint on the edge.

 We call the maximum amount by which we can increase the flow on each edge in an augmenting path p, the residual capacity of p, given by $c_f$(p) :

$$c_f(p) = \min \{c_f(u, v) : (u, v) \text{ is on } p\} \qquad (3)$$

*Cuts of Flow Networks:* The Ford-Fulkerson method repeatedly augments the flow along augmenting paths until a maximum flow has been found. The max-flow min-cut theorem, tells us that a flow is maximum if and only if its residual network contains no augmenting path.

*The Ford-Fulkerson Method*: The Ford-Fulkerson method is iterative. We start with $f(u, v) = 0$ for all $u, v \in V$, giving an initial flow of value 0. Each iteration, we increase the flow value by finding an augmenting path, which we can think of simply as a path from the source s to the sink t along which we can send more flow, and then augmenting the flow along this path. We repeat this process until no augmenting path can be found. The max-flow theorem shows that upon termination, this process gives a maximum flow.

Each iteration of the Ford-Fulkerson method, we find some augmenting path p and increase the flow f on each edge of p by the residual capacity $c_f(p)$. The following implementation of the method computes the maximum flow in a graph $G = (V, E)$ by updating the flow $f[u, v]$ between each pair $u, v$ of vertices that are connected by an edge. If u and v are not connected by an edge in either direction, we assume implicitly that $f[u, v] = 0$. The capacities $c(u, v)$ are assumed to be given along with the graph, and $c(u, v) = 0$ if $(u, v)$ is not in E. The residual capacity $c_f(u, v)$ is computed in accordance with the formula (1). The expression $c_f(p)$ in the code is actually just a temporary variable that stores the residual capacity of the path p.

*Maximum Bipartite Matching:* The problem of finding a maximum matching in a bipartite graph can be reused to maximum flow problem. It means that if we can solve the maximum flow problem, we have solved the maximum matching problem.

Therefore, if the Ford-Fulkerson can solve the maximum flow problem, it can solve the problem of maximum matching in a bipartite graph, too.

---

**ALGORITHM 2 : Ford-Fulkerson-Method (G, s, t)**

*1 :*   *initialize flow f to 0*
*2 :*   *while there exists an augmenting path p*
*3 :*      *do augment flow f along p*
*4 :*   *return f*

---

**ALGORITHM 3 : Ford-Fulkerson (G, s, t)**

*1 :*   *for each edge (u, v) $\in E[G]$*
*2 :*   *do f [u, v] ← 0*
*3 :*    *f [v, u] ← 0*
*4 :*   *while there exists a path p from s to t in the residual network $G_f$*
*5 :*   *do $c_f(p)$ ← min $\{c_f(u, v) : (u, v)$ is in p$\}$*
*6 :*    *for each edge (u, v) in p*

> *7 :*        *do f [u, v] ← f [u, v] + c_f(p)*
> *8 :*            *f [v, u] ← − f [u, v]*

*The Maximum Bipartite-Matching Problem*: Given an undirected graph G = (V, E), a matching is a subset of edges M ⊆ E such that for all vertices v ∈ V, at most one edge of M is incident on v. We say that a vertex  v ∈ V is matched by matching M if some edge in M is incident on v; otherwise, v is unmatched. A maximum matching is a matching of maximum cardinality, that is, a matching M such that for any matching M′, we have |M| ≥ |M′|.

We assume that the vertex set can be partitioned into V = L ∪ R, where L and R are disjoint and all edges in E go between L and R. We further assume that every vertex in V has at least one incident edge. We have shown a bipartite graph in "figure 2".

*Finding A Maximum Bipartite Matching*: We can use the Ford-Fulkerson method to find a maximum matching in an undirected bipartite graph G = (V, E) in time polynomial in |V| and |E|. The trick is to construct a flow network in which flows correspond to matching, we define the corresponding flow network G′ = (V′, E′) for the bipartite graph G as follows. We let the source s and sink t be new vertices not in V, and we let V′ = V ∪ {s, t}. If the vertex partition of G is V = L ∪ R, the directed edges of G′ are the edges of E, directed from L to R, along with V new edges:

E′ ={(s, u) : u ∈ L}∪{(u, v) : u ∈ L, v ∈ R, and (u,v) ∈ E}∪{(v,t):v ∈ R}    (4)

To complete the construction, we assign unit capacity to each edge in E′.

In "figure 3", we have shown a bipartite graph which is made up of output parameters of requested (A) and advertised (B) web services and augmenting path is shown by dark lines.

By applying Ford Fulkerson algorithm on this bipartite graph, we can compute the matching rate between two web services, A and B. we should make such a graph for these two services` inputs ($A_{in}$, $B_{in}$). We consider the average of input similarity rate and output similarity rate for two web services similarity rate.

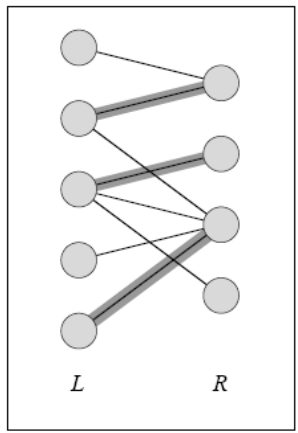

**"Figure 2. A bipartite graph"**

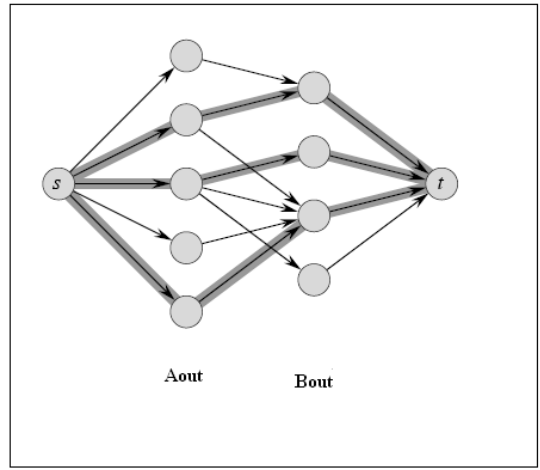

**"Figure 3. Bipartite matching"**

**3.1.2. Our matching algorithm:** If the result of matching two services` output parameters is OUTSIM and the result of matching two services` input parameters is INSIM, the whole result of matching two services is PARSIM that obtains by the following algorithm. The input of our algorithm is B which is the matched service. The output of our algorithm is PARSIM. It has a sharp result because we always choose the lowest matching rate. So if we had 10 as the result, it means that two services match completely.

---

**Algorithm 4: Parameter Match (B, PARSIM);**

*1 :  for all output parameters of A do*
*2 :     Case ($A_{out}$, $B_{out}$ ) ;*
*3 :     $G_{out}$= make a bipartite graph for outputs;*
*4 :     OUTSIM = Ford-Fulkerson ($G_{out}$);*
*5 :  for all input parameters of A do*
*6 :     Case (Ain, $B_{in}$);*
*7 :     $G_{in}$= make a bipartite graph for inputs;*
*8 :     INSIM= Ford-Fulkerson ($G_{in}$);*
*9 :  PARSIM = 10 ;*
*10 :if (OUTSIM=0 or INSIM=0 ) then*
*11 :   PARSIM = 0*
*12 :   else if (OUTSIM=3 or INSIM=3 ) then*
*13 :      PARSIM = 3*
*14 :      else if (OUTSIM=7 or INSIM=7 ) then*
*15 :         PARSIM = 7 ;*

### 3.2. Second phase

In this phase we compare the type of parameters of input and output. At first we should apply the rules and then we should make a bipartite graph for both input and output parameters. The vertices are the input/output parameters. The weights of edges are according to the "table 1". After making a weighted bipartite graph, by the help of Ford-Fulkerson algorithm, compute the type matching rate. The output of this part is TYPESIM.

**"Table 1. Rules of comparing two parameters of two web services"**

| | | B Parameter | | | | |
|---|---|---|---|---|---|---|
| | *Data Type* | Integer | Real | String | Date | Boolean |
| **A Parameter** | Integer | 10 | 5 | 3 | 1 | 1 |
| | Real | 10 | 10 | 1 | 0 | 1 |
| | String | 7 | 7 | 10 | 8 | 3 |
| | Date | 1 | 0 | 1 | 10 | 0 |
| | Boolean | 1 | 0 | 1 | 0 | 10 |

### 3.3. Third phase

Now we have the results of first two phases, PARSIM and TYPESIM. It is obvious that semantics of the parameters have the main role of matchmaking and semantic meaning of a parameter is more important than its type. So we compute the final result according to algorithm 5:

```
Algorithm 5: Final;

1 :  If PARSIM=0 then
2 :     result = 0
3 :     else
4 :        result=[2/3(PARSIM) + 1/3(TYPESIM)]*100
```

## 4. Evaluation of work

One of the best ways for evaluating a solution is computing the running time of the method. Our matchmaking is composed of three phases. The time order of two first phases is the same because their solutions are the same. The time for making a bipartite graph and put the weights on the edges, according to the rules, is done in polynomial time. But the bipartite matching time which is done by the Ford-Fulkerson algorithm, should be computed.

The running time of Ford-Fulkerson depends on how the augmenting path p in line 4 of algorithm 2 is determined. If the augmenting path is chosen by using a breadth-first search, the algorithm runs in polynomial time. Augmenting path is chosen arbitrarily and all capacities are integers which are in our usage, between 0 and 10. A straightforward implementation of Ford-Fulkerson runs in time $O(E \mid f_{max} \mid)$, where $f_{max}$ is the maximum flow found by the algorithm. The analysis is as follows.

Lines 1–3 take time 2(E). The while loop of lines 4–8 is executed at most $|f_{max}|$ times, since the flow value increases by at least one unit in each iteration. The work done within the while loop can be made efficient if we efficiently manage the data structure used to implement the network G = (V, E). Let us assume that we keep a data structure corresponding to a directed graph G′ = (V, E′), where E′= {(u, v): (u, v) ∈ E or (v, u) ∈ E}. Edges in the network G are also edges in G′, and it is therefore a simple matter to maintain capacities and flows in this data structure. Given a flow f on G, the edges in the residual network $G_f$ consist of all edges (u, v) of G′ such that c (u, v) − f [u, v] ≠ 0. The time to find a path in a residual network is therefore O (V + E′) = O (E) if we use either depth-first search or breadth-first search. Each iteration of the while loop thus takes O(E) time, making the total running time of Ford-Fulkerson O(E $|f_{max}|$).

As you see in "table 2" all other algorithms that compute the bipartite matching have more running time than Ford-Fulkerson.

**"Table 2. Comparision of different algorithms` running time"**

| Algorithm `s name | Running time |
|---|---|
| Ford-Fulkerson | O(E $|f_{max}|$) |
| Edmonds-Karp | O(V.E$^2$) |
| Relabel-To-Front | O(V$^3$) |
| Push-Relabel | O(V$^2$E) |

The advantages of our proposed approach compared to the other approaches can be categorized as following:

- We have laid the foundation of our approach on the top of semantic Web standards: using the semantic matching and the Semantic annotations for Web services and Request description, comparing to the syntactic methods;

- The exploration in a single pass (according to the Ford-Fulkerson algorithm) over the flow network reduces again the time response comparing to the other methods [17,18];

- Our technique aims to reduce firstly, the complexity of the matchmaking and secondly, the time needed to response the requester by selecting the best, similarity measure and exec-time ,at run time automatically, comparing to this approach [22] and others;

- In our approach, we are not obliged to know the behavior of a Web service compared to the methods [5], the behavior is a complex feature in matchmaking task;

- Our prototype was tested on existing services comparing to most approaches which do not give any details of implementation.

## 5. Conclusion and future work

Web services should be described in a way that end user can use it simply. The first choice that comes into mind is Web Service Description Language (WSDL). In theory, semantic information in WSDL files was supposed to solve this problem, because WSDL is a way to know what a service does and how. But in practice, is not enough, because currently WSDL files don't have enough semantic information to decide substitutability or ability of compose. There is a need for automatic techniques to obtain more semantic information. One of the best ways to come over this problem is using an Ontology Web Language, which has the meaning of different parameters of a service.

In this paper we present a novel approach for semantic web matching. Our matchmaker system has two phases for semantic web matching and in both phases, it uses bipartite graph for computing the matching rate. The innovation of this work is using flow networks for computing matching rate. We also presented a scoring rule to be used as the weights of the bipartite graph. By the help of Ford-Fulkerson algorithm, that is the best algorithm among the ones who are used for bipartite matching. another part of this work`s innovation is matchmaking both semantic and type of the parameters. We described our services all in the same Ontology Web Language. So our comparison of web services was in a standard way.

The ideas given in this paper also leave open some other interesting research issues of Web Services from the practical point of view. Other than precise search and automatic composition of Web services, our methods can be extend to many related real applications of domain-specified services management, such as decision-making, prediction, trend analysis, and so on.

In addition with input and output parameters, precondition and event are two another factors for comparing semantic web matching.

By considering QoS factors like response time, we can give a better answer to our service requesters. So our future work focuses on comparing preconditions and events of two web services and considering agent requests and QoS.

# Authors

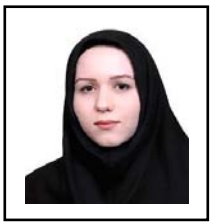

**Golsa Heidary** is a researcher and a member of Young Researchers Club, Department of Computer, Islamic Azad University of Najafabad. She received her BSc degree from Univ. of Isfahan, M.Sc. in software engineering from the Faculty of Computer, Islamic Azad University of Najafabad, Isfahan, Iran (2010). Her main interests include distributed systems and web service technology. She has published some conferences papers.

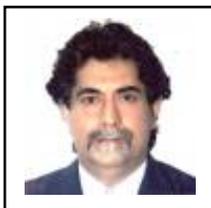

**Kamran Zamanifar** is Assistant Professor at the University of Isfahan since 1996. He received his M.Sc. in electrical and electronic engineering from the Faculty of Engineering, University of Tehran, Iran. He also received his Ph.D. in computer science (parallel and distributed systems) from the School of Computer Studies, University of Leeds, England (1992-1996). He is a member of various scientific committees such as Management Board of Computer Society of Iran, Iranian Association of Electrical and Electronic Engineers and Annual International CSI

Computer Conferences. He is also reviewer of many scientific journals. His main interests include parallel and distributed systems, concurrent systems and high performance computing. He has published two books and some journals and conferences papers.

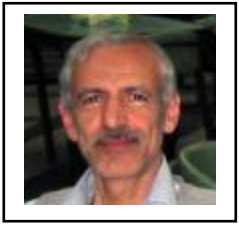 **Naser Nematbakhsh** received his BSc degree from Univ. of Isfahan 1973, MS degree from Worcwstechnic Inst., USA in 1978 and PhD from Bradford Univ. UK in 1989. He is now assistant professor in computer engineering dept. at University of Isfahan. His main research interest include Modeling, Perfeormance evaluation, Reliability and software engineering.